\documentclass[preprint]{elsarticle} %alternative preprint or review

\usepackage{lineno,hyperref}
\usepackage{ragged2e}
\usepackage[format=plain,justification=RaggedRight,  singlelinecheck=false]{caption}
          
\usepackage{siunitx}         % set units as follows: \SI{3}{mm}
\usepackage{float}	         % use [H] to force figure here
\usepackage{threeparttable}  % create tables with tablenotes
\usepackage{multirow}        % use multirow in threeparttable
\usepackage{bm}				 % use bold font
\usepackage{amsmath}		 % use \eqref to cite equations
\usepackage{subcaption}		 % substructure of captions and figures
\usepackage{fixltx2e} 		 % subscript in Text mode
\usepackage{pdfpages} 		 % PDF einfügen
\usepackage{subcaption}		 % subcaptions under part images
\graphicspath{{images/}}

\usepackage{booktabs}

\usepackage{booktabs}
\usepackage{longtable}

\usepackage{siunitx}
\usepackage{nomencl}
\usepackage{ifthen}
\renewcommand\nomgroup[1]{%
  \ifthenelse{\equal{#1}{A}}{%
    \item[\textbf{Acronyms}]}{%                A - Acronyms
  \ifthenelse{\equal{#1}{R}}{%
    \item[\textbf{Roman Symbols}]}{%           R - Roman
  \ifthenelse{\equal{#1}{G}}{%
    \item[\textbf{Greek Symbols}]}{%           G - Greek
  \ifthenelse{\equal{#1}{S}}{%
    \item[\textbf{Superscripts}]}{%            S - Superscripts
  \ifthenelse{\equal{#1}{U}}{%
    \item[\textbf{Subscripts}]}{%              U - Subscripts
  \ifthenelse{\equal{#1}{X}}{%
    \item[\textbf{Other Symbols}]}{%           X - Other Symbols
  {}}}}}}}}

\journal{arXiv} % add journal name

\begin{document}

\begin{frontmatter}

%\tableofcontents			 % add table of contents

\title{Validation of a temperature-dependent elasto-viscoplastic material model for a talcum-filled polypropylene/polyethylene co-polymer using glove box flap component tests}
% \tnoteref{mytitlenote}}

\author{D. Degenhardt\fnref{mymainaddress,mysecondaryaddress}}\ead{david.degenhardt@volkswagen.de}
\author{J. Langer\fnref{mythirdaddress}}
\author{L. Greve\fnref{mymainaddress}}
\author{T.K. Eller\fnref{mymainaddress}}
\author{M. Andres\fnref{mymainaddress}}
\author{P. Horst\fnref{mysecondaryaddress}}

\address[mymainaddress]{Volkswagen AG, Group Research, P.O. Box 1777, 38436 Wolfsburg, Germany}
\address[mysecondaryaddress]{TU Braunschweig, IFL, Herrmann-Blenk-Str. 35, 38108 Braunschweig, Germany}
\address[mythirdaddress]{Leibniz Universit\"{a}t Hannover, IKM, Welfengarten 1, 30167 Hannover, Germany}

\begin{abstract}
In the automotive industry, thermoplastic polymers are used for a significant number of interior and exterior parts. These components have to pass all underlying crash and safety relevant tests, where a proper performance is desired in the range of low to high ambient temperatures. Today, the vehicle design is heavily aided by numerical simulation methods for advancing towards a prototype free vehicle development. This requires an accurate modeling of the temperature- and rate-dependent, elasto-viscoplastic mechanical response of the polymer structures. In this work, the validation of a novel elasto-viscoplastic temperature-dependent material model is performed using glove box flap segments subjected to impact loading by a spherical punch in a custom-build loading frame. The proposed material model shows a very good prediction of the experimental results. \\
\end{abstract}

\begin{keyword}
temperature-dependency, fracture modeling, polymer, validation, component testing
%%%\texttt{elsarticle.cls}\sep \LaTeX\sep Elsevier \sep template
%%%\MSC[2010] 00-01\sep  99-00
\end{keyword}

\end{frontmatter}

%%%\newpage

%%%\noindent \textbf{Scope of the Journal Polymer Testing} \\

%%%\noindent Polymer Testing focuses on the testing, analysis and characterization of polymer materials, including both synthetic and natural or biobased polymers. Novel testing methods and the testing of novel polymeric materials in bulk, solution and dispersion is covered. In addition, we welcome the submission of the testing of polymeric materials for a wide range of applications and industrial products as well as nanoscale characterization. \\

%%%\noindent The scope includes but is not limited to the following main topics: \\

%%%\begin{itemize}
%%%\item Novel testing methods and chemical analysis (mechanical, thermal, electrical, chemical, imaging, spectroscopy, scattering and rheology)
%%%\item Physical properties and behaviour of novel polymer systems (nanoscale properties, morphology, transport properties)
%%%\item Degradation and recycling of polymeric materials when combined with novel testing or characterization methods (degradation, biodegradation, ageing and fire retardancy)
%%%\end{itemize}

%%%\noindent Modelling and Simulation work will be only considered when it is linked to new or previously published experimental results.

\newpage

\section*{Nomenclature}

\subsection*{Abbreviations}
\markboth{Nomenclature}{Abbreviations}

\begin{longtable}{p{0.2\textwidth} p{0.75\textwidth}}
  \toprule
  Abbreviation & Meaning \\
  \midrule
  BT & \underline{b}i-axial \underline{t}ension\\
  FE & \underline{f}inite \underline{e}lement \\
  HA & \underline{Ha}sek \\
  MMM & \underline{m}odular \underline{m}aterial \underline{m}odel \\
  PE & \underline{p}oly\underline{e}thylene \\
  PP & \underline{p}oly\underline{p}ropylene \\
  PT & \underline{p}unch \underline{t}est \\
  PTFE & \underline{p}oly\underline{t}etra\underline{f}luoro\underline{e}thylene \\
  RT & \underline{r}oom \underline{t}emperature \\
  SH & \underline{sh}ear \\
  UC & \underline{u}ni-axial \underline{c}ompression \\
  UT & \underline{u}ni-axial \underline{t}ension \\
  
  \bottomrule
\end{longtable}

\subsection*{Latin designations}
\begin{longtable}{p{0.1\textwidth} p{0.2\textwidth} p{0.65\textwidth}@{}}
  \toprule
  Symbol & Dimension & Description\\
  \midrule

  $c$ & --- & fracture or hardening parameter \\
  $d$ & \SI{}{mm} & displacement \\
  $f$ & --- & scaling factor \\
  $g$ & \SI{}{MPa} & plastic potential \\
  $k$ & --- & strain rate hardening function \\
  $l$ & \SI{}{mm} & length \\
  $t$ & \SI{}{ms} & time \\
  $v$ & \SI{}{mm\,ms\textsuperscript{-1}} & velocity \\
  $w$ & \SI{}{mm} & width \\
   \  &     \      &        \        \\

  $E$ & \SI{}{MPa} & elastic modulus \\
  $F$ & \SI{}{N} & force \\  
  $R$ & \SI{}{GPa} or \SI{}{mm} & hardening function or radius \\  
  $T$ & \SI{}{^{\circ} C} & temperature \\  
   \  &     \      &        \        \\
  FRAC & --- & maximum accumulated damage parameter \\    
  \bottomrule
\end{longtable}

\subsection*{Greek designations}
\begin{longtable}{p{0.1\textwidth} p{0.2\textwidth} p{0.65\textwidth}@{}}
  \toprule
  Symbol & Dimension & Description\\
  \midrule
  $\gamma$ & --- & yield function parameter \\  
  $\delta$ & --- & flow rule parameter \\
  $\varepsilon$ & --- & true strain \\  
  $\dot{\varepsilon}$ & \SI{}{s^{-1}} & strain rate \\ 
  $\bar{\varepsilon}$ & --- & equivalent strain \\ 
  $\eta$ & --- & stress triaxiality \\   
  $\theta$ & --- & Lode angle \\  
  $\bar{\theta}$ & --- & normalized Lode angle \\
  $\nu$ & --- & Poisson's ratio \\        
  $\sigma$ & \SI{}{MPa} & true stress or yield strength \\
  $\bar{\sigma}$ & \SI{}{MPa} & equivalent stress \\
  $\bm{\sigma}$ & \SI{}{MPa} & Cauchy stress tensor \\ 
  
    \bottomrule
\end{longtable}

\subsection*{Subscripts and superscripts}
\begin{longtable}{p{0.1\textwidth} p{0.2\textwidth} p{0.65\textwidth}@{}}
  \toprule
  Symbol & Dimension & Description\\
  \midrule
  $x^\text{e}$ & --- & elastic \\
  $x^\text{p}$ & --- & plastic \\
       \       &  \  &   \     \\
   
  $x_\text{c}$ & --- & compressive \\
  $x_\text{f}$ & --- & fracture \\
  $x_\text{m}$ & --- & hydrostatic \\
  $x_\text{qs}$ & --- & quasi-static \\
  $x_\text{t}$ & --- & tensile \\
  $x_\text{vM}$ & --- & von Mises \\
           
  \bottomrule
\end{longtable}

\subsection*{Symbols and operators}
\begin{longtable}{p{0.1\textwidth} p{0.2\textwidth} p{0.65\textwidth}@{}}
  \toprule
  Symbol & Dimension & Description\\
  \midrule
  $\Delta$ & --- & differential operator\\
  \bottomrule
\end{longtable}

\newpage

%%%%%\linenumbers

\section{Introduction}

\noindent Polymers in the automotive industry are attractive as they can be used for modeling of complex structures while being cost-efficient \cite{Maddah2016}. For the prediction of the material behavior of unstiffened thermoplastic polymers, there are several features that have to be accounted for. Such properties include, among others, the rate- and temperature-dependency of the mechanical properties, the yield strength difference in tension and compression and the plastic softening driven by volume dilatation \cite{Hall1968, Boyce1989, Santangelo1996, Zhou2002}. \\

\noindent A phenomenological temperature-dependent material model capturing the aforementioned polymer-specific phenomena was developed by Degenhardt et al.~\cite{Degenhardt2019}. It relies upon a profound material characterization at the three main supporting temperatures $\SI{-35}{^\circ \text{C}}$, $\SI{20}{^\circ \text{C}}$ and $\SI{90}{^\circ \text{C}}$, considering several stress states such as uni-axial tension (UT), uni-axial compression (UC), bi-axial tension (BT) and shear (SH). The temperature-dependent material model is built on the basis of the modular material model (MMM) developed by the Volkswagen Group Research, which has been previously used for modeling polymers and several metals \cite{Greve2018, Greve2012, Eller2014, Greve2016, Eller2017}. As a dynamic UserMaterial library, the MMM is linked to the explicit crash code Virtual Performance Solution \cite{VPS2017}. \\

\noindent In this work, the temperature-dependent material model is validated using a component test: punch test with a glove box flap component. The material model is validated at a combination of stress states and ambient temperatures which were not used for model calibration. The strategy for validating the material model reads as follows:

\begin{itemize}
\item Comparison of material properties of UT specimens extracted from the glove box flap component and from the plates used for calibration, revealing some differences of the mechanical properties with regard to the fracture behavior. These effects will be taken into account by adjustment of the fracture model.
\item Execution of punch tests with glove box flap components at different ambient temperatures
\item Conduction of finite element (FE) simulations of the component impact test
\item Comparison of the force-displacement curves as well as the initial fracture positions of the physical experiments and the corresponding simulations
\end{itemize}

\section{Material description} \label{Ch:material_description}

\noindent In this work, the investigated glove box flap component is made of a $\SI{20}{\text{wt}\%}$ talcum-filled semi-crystalline polypropylene (PP) / polyethylene (PE) co-polymer. The use of the talcum filler aims to reduce costs while increasing crystallinity and stiffness as well as decreasing the tenacity and impact strength  \cite{Ammar2017}. The blend is widely used in the automotive industry for several interior parts as well as paneling. Consisting of little quantities of PE ($\leq \SI{5}{\text{wt}\%}$), the blend behavior is estimated to be similar to a pure PP but with increased ductility because of the weaker PE. \\

\begin{figure}[H]
\centering
\includegraphics[width=\textwidth]{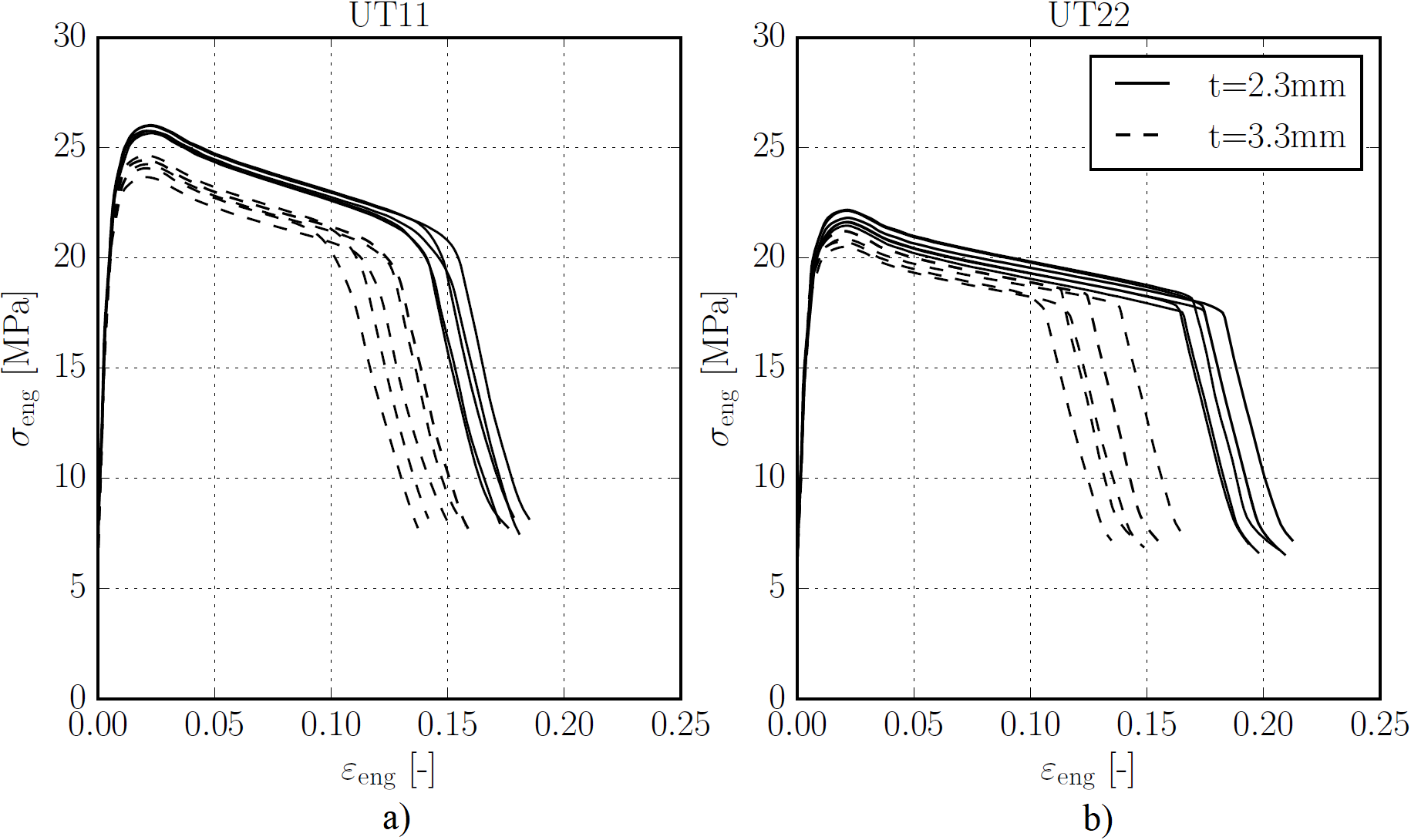}
\caption{Thickness study of the talcum-filled PP/PE co-polymer for specimens in (UT11, a) and perpendicular to (UT22, b) the mold flow direction}
\label{fig:thickness_study}
\end{figure}

\noindent Depending on the manufacturing process, unstiffened polymers may also exhibit anisotropic behavior and thickness sensitivity \cite{Hnatkova2016}. The reason for the anisotropic behavior is that the polymer chains have a high degree of molecular orientation in the mold flow direction. Moreover, preliminary experiments have shown that the strength loss between specimens tested along (UT11) and perpendicular (UT22) to the mold flow direction is around 15\%, compare Figures~\ref{fig:thickness_study}a and \ref{fig:thickness_study}b. For a more conservative material design, the UT22 material properties are used for model calibration. The loss in fracture strain due to the thickness effect from $\SI{2.3}{mm}$ to $\SI{3.3}{mm}$ is around 30\% for UT22 at room temperature. 

\noindent The thickness sensitivity, also illustrated in Figures~\ref{fig:thickness_study}a and \ref{fig:thickness_study}b, comes into place because of the skin-core effect. In 1972 Kantz et al.~\cite{Kantz1972} first saw that the skin-core morphology of their injection-moulded PP samples is influenced by molecular properties such as the molecular weight, and also by the processing conditions. They showed that the melt temperature affects the crystallite orientation which is the major parameter for several material properties, such as the tensile and impact strength or the shrinkage. The observations were later confirmed by Altendorfer and Seitl in 1986 \cite{Altendorfer1986}, who focused on the influence of the molecular weight and the molecular weight distribution of injection-moulded samples. They varied the molecular weight and the molecular weight distribution of their PP samples and concluded that the skin-core morphology varies depending not only on the flow direction but also depending on the molecular characteristics. Consequently, the skin-core effect causes the polymer chains in the skin zone to have a high degree of molecular orientation in the flow direction of the melt due to the molecular motions forming new crystal structures. This results in a higher strength under tensile load in this direction and after Karger-Kocsis and Friedrich also in a higher resistance against fatigue rupture \cite{Karger1989}. \\

\noindent As a result, when increasing the specimen thickness and assuming a constant width of the skin layers, the ratio of skin to core zones decreases, leading to a lower overall strength. Therefore,
polymers may diverge from the assumed isotropic material behavior, depending on their flow direction, flow velocity and cooling rate from the melt as well as their thickness. \\

\section{Brief review of the material model}

\begin{figure}[H]
\centering
\graphicspath{{images/}}
\def\svgwidth{1.0\textwidth}
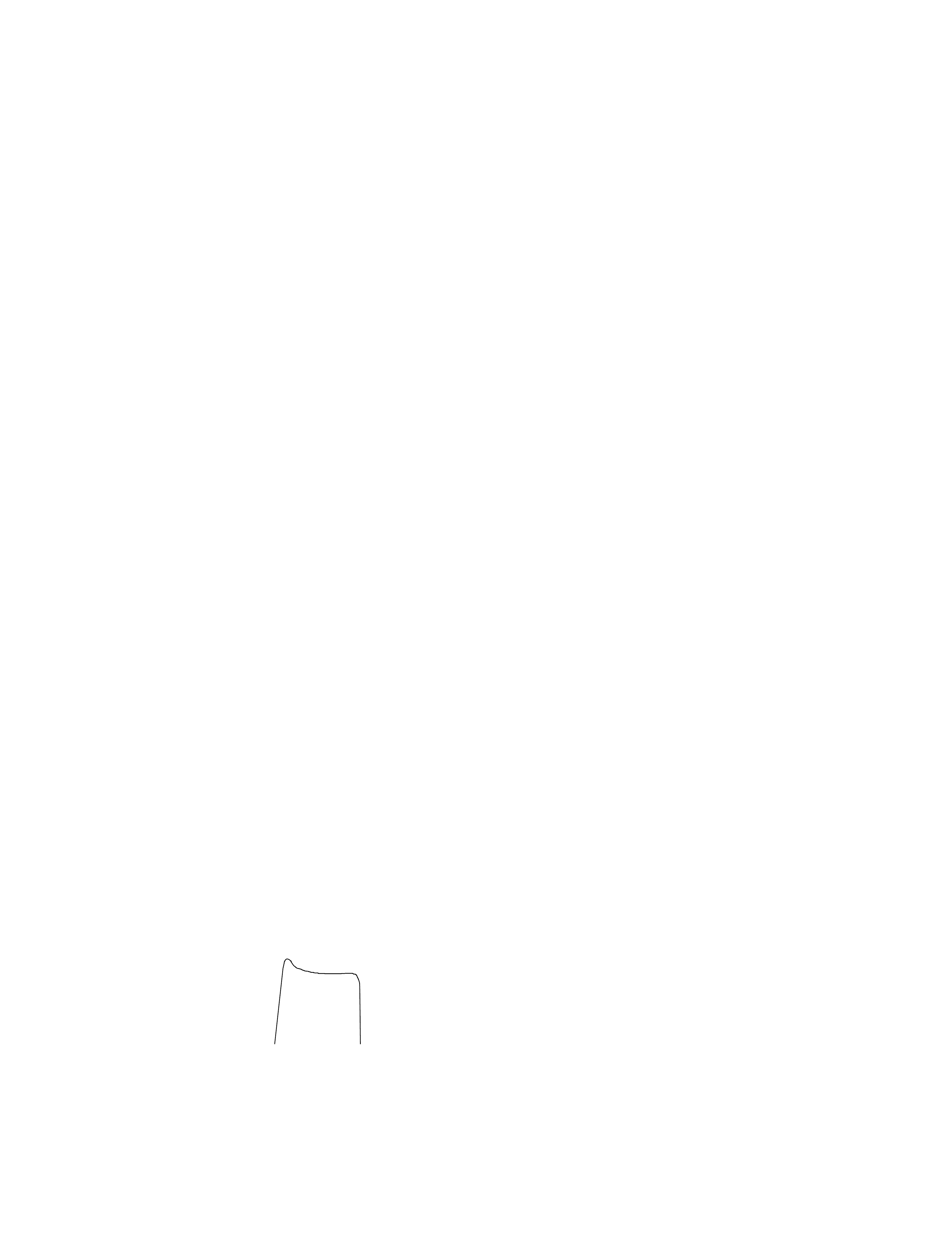
\caption{Review of the numerical model from \cite{Degenhardt2019}}
\label{fig:Review_model}
\end{figure}

\noindent The temperature-dependent material model proposed in \cite{Degenhardt2019} is a generalized constitutive model representing the deformation and fracture behavior of the PP/PE co-polymer mentioned above for the crash-relevant temperature range from $\SI{-35}{^\circ \text{C}}$ to $\SI{90}{^\circ \text{C}}$. For capturing the temperature-dependency, a non-linear interpolation concept is introduced. The philosophy of the material model development is briefly explained in the following, from material characterization to parameter identification to model/test comparison. An overview of the process steps is given in Figure~\ref{fig:Review_model}. Generally, the material model is calibrated in two steps. \\

\noindent In a first step, a profound material characterization is performed at the three main supporting temperatures $\SI{-35}{^\circ \text{C}}$, $\SI{20}{^\circ \text{C}}$ and $\SI{90}{^\circ \text{C}}$. The material tests UT, UC, SH, BT and plane strain tension using the Hasek (HA) test \cite{Hasek1978} are conducted. During the experiments, the global measurements are the force and the displacement. In the UT tests, the local strain field is measured in addition. This is highly recommended because the fracture criterion is calibrated using the strain field as foundation. The optimal parameters for the material model are identified by a minimization of target functions considering both the global forces and displacements as well as the local strains. Consequently, the optimal parameter set often tends to be a trade-off, however, this way the best fit of both quantities is met, resulting in a well-founded set of parameters. For the material model calibration a total of 35 modeling parameters per main supporting ambient temperature are identified. This includes, among others, the parameters for the linear isotropic elasticity $E$ and $\nu^\text{e}$, for the extended Raghava isotropic yield function $\bar{\sigma}(\bm{\sigma})$ with non-associated parabolic flow rule $g(\bm{\sigma})$, for the analytical hardening law $R_\text{qs}$ with non-linear rate-dependent term $k_{\dot{\varepsilon}}$ as well as parameters for the rate-dependent fracture strain $\bar{\varepsilon}_\text{f}$, depending on the normalized Lode angle parameter $\bar{\theta}$ and the stress triaxiality $\eta$. \\

\noindent In a second step, the non-linear interpolation for the temperature-dependency is calibrated with additional UT tests at seven ambient temperatures between the main supporting temperatures. From the additional UT tests alone, a majority of parameters are provided including the elastic modulus $E$, the non-linear plastic Poisson's ratio $\nu^\text{p}$, the hardening function $R$ and the equivalent fracture strain $\bar{\varepsilon}_\text{f}$ in UT. The remaining parameters, which are solely available at the main supporting temperatures, are interpolated linearly. \\

\noindent In the material model, the parameter FRAC is defined as maximum accumulated damage of each finite element. The calculation of the fracture indicator is given in Equation~\eqref{eq:FRAC_MMC} with more details in \cite{Greve2018}. An undamaged material has a FRAC-value of 0. At FRAC$=$1 fracture initiates, triggering element elimination in the FE-model.

\begin{equation} \label{eq:FRAC_MMC}
\text{FRAC} = \int \! \frac{\Delta \bar{\varepsilon}^\text{p}}{\bar{\varepsilon}_\text{f} \left( \eta, \bar{\theta}, c_\text{1-4}, \dot{\bar{\varepsilon}}^\text{p} \right)}
\end{equation}

\section{Experimental work} \label{Sec:experimental_work}

\subsection{Overview}

\begin{figure}[H]
\centering
\graphicspath{{images/}}
\def\svgwidth{1.0\textwidth}
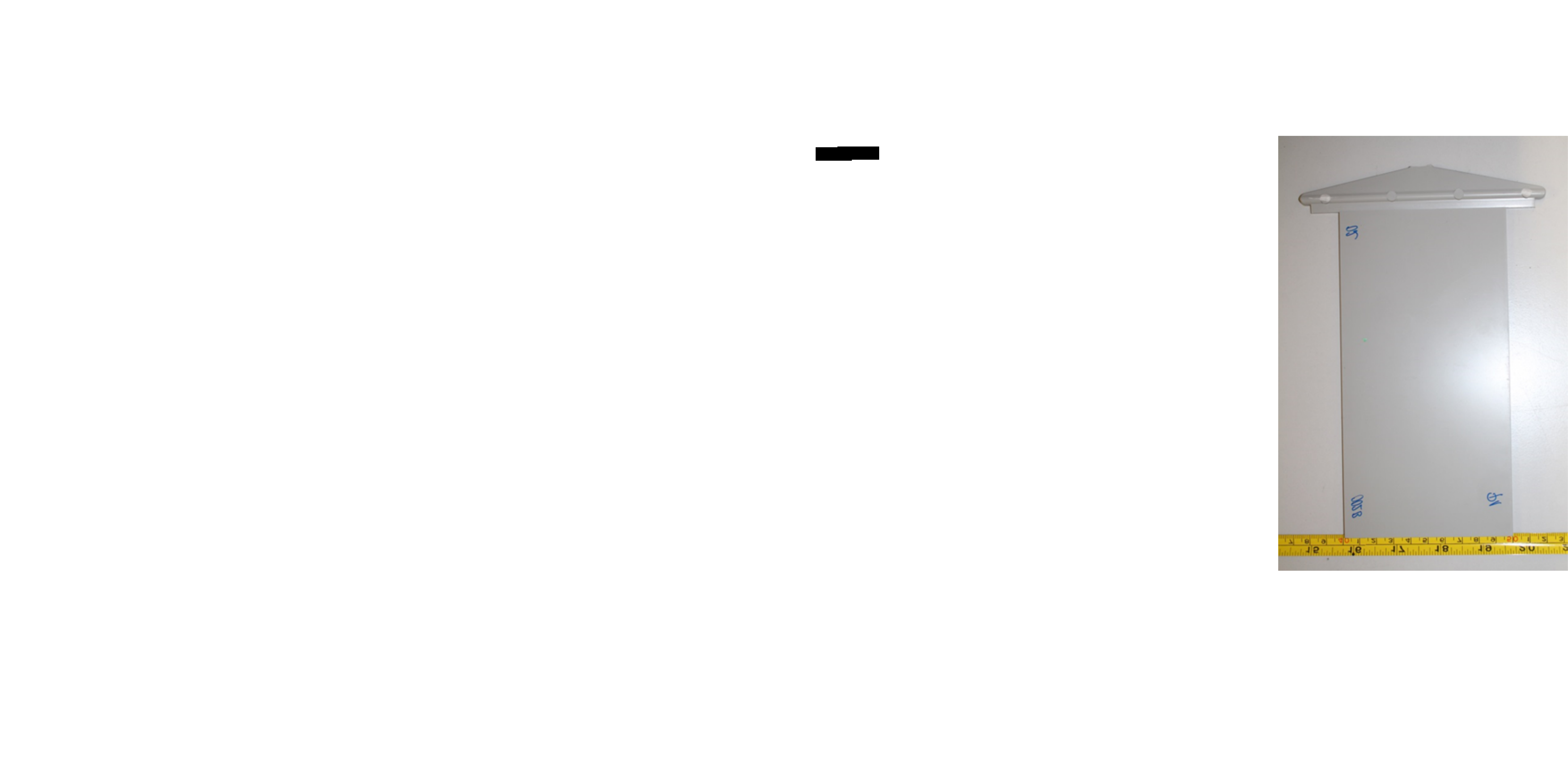
\caption{Extraction of a) cutouts and UT specimens out of a glove box flap and b) UT specimens out of a plate used for material characterization}
\label{fig:component_cutouts}
\end{figure}

\noindent For the validation of the temperature-dependent material model, an idealized knee-impact of a glove box flap, Figure~\ref{fig:component_cutouts}a, is investigated. From this component, the two highlighted cutouts, the glove box flap and the UT specimen, are extracted by milling. The glove box flap cutouts are used for punch tests (PT). The shape of this part is large enough to represent the component complexity while it is small enough to fit into a temperature chamber during testing. The UT specimen is extracted from an area with constant thickness and without reinforcing ribs. It is taken from the component in order to check whether the material properties of the components match the material behavior of UT specimens from the plates, which were used for model calibration, Figure~\ref{fig:component_cutouts}b. The temperature-dependent material model was calibrated using UT specimens extracted from $\SI{2.3}{mm}$ thick plates with a more than five times wider sprue gate than in the component. Both sprue gate lengths are marked in Figure~\ref{fig:component_cutouts}. The tested configurations are summarized in Table~\ref{tab:testing_conditions}. \\

\begin{table}[H]
\caption{Tested configurations}
\centering
\begin{tabular}{cccccc}\toprule
Test type & Label & Test velocity [mm\,s\textsuperscript{-1}] & \multicolumn{3}{c}{Temperature [$^\circ \text{C}$]}  \\ \midrule
uni-axial tension & UT & 1.4 & -35,\;\; &  20, & 90 \\
punch test        & PT & 1.0 & -10,\;\; &  23, & 50 \\ \bottomrule
\end{tabular}
\label{tab:testing_conditions}
\end{table}

\subsection{Uni-axial tension tests}

\begin{figure}[H]
\centering
\begin{minipage}{1.0\textwidth}
\includegraphics[width=\textwidth]{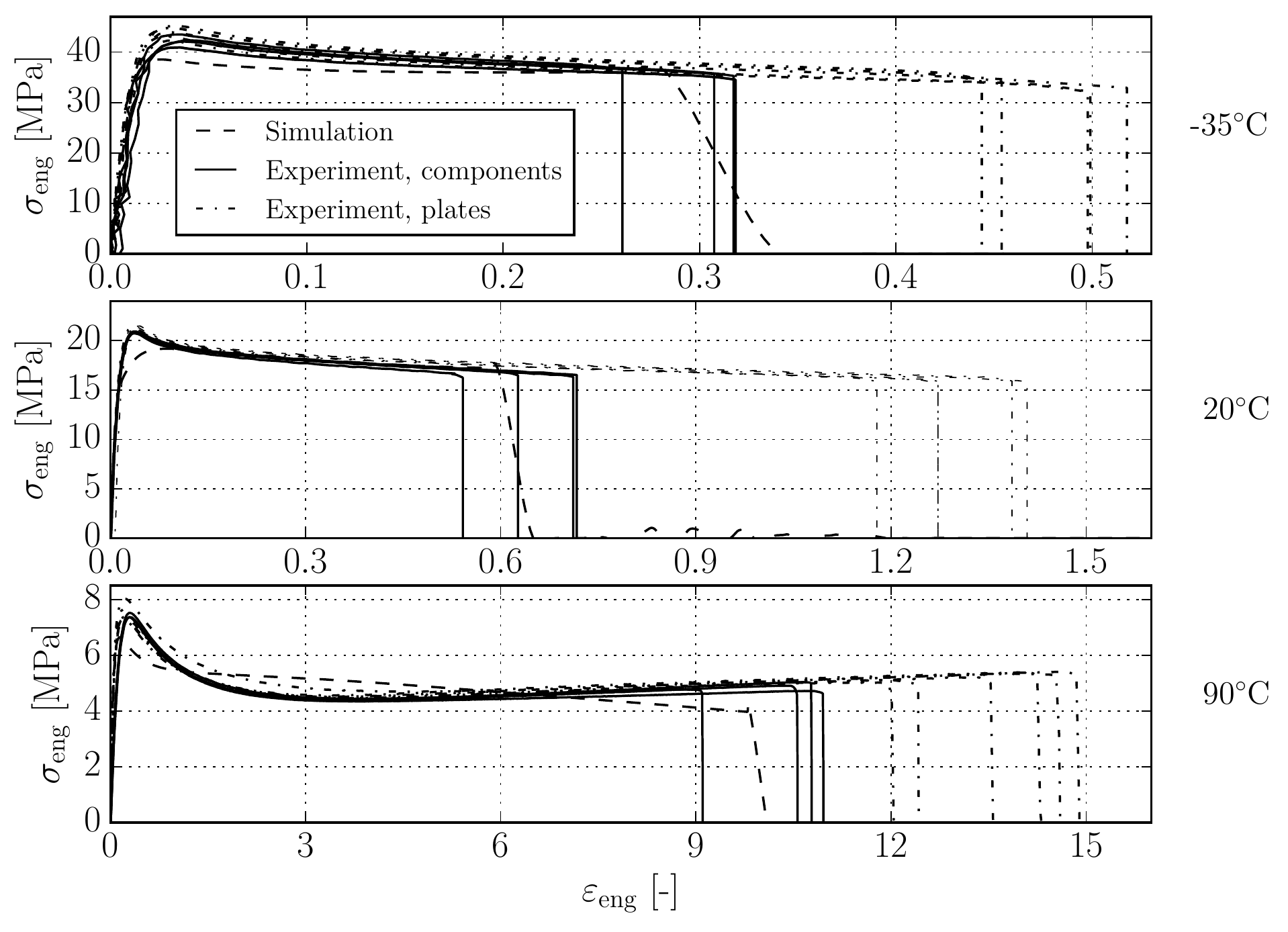}
\end{minipage}
\hfill
\caption{Comparison of the simulation and the experimental data of the UT tests with specimens from plates and from components}
\label{fig:UTStressStrainDINIT}
\end{figure}

\noindent For the tensile test component cutouts, the UT cubic spline specimen geometry from \cite{Degenhardt2019} was used. The experimental results from the component UT cutouts as well as from the calibration test from plates are given in engineering stress-strain diagrams in Figure~\ref{fig:UTStressStrainDINIT} for the main supporting temperatures.

\begin{figure}[H]
\centering
\includegraphics[width=\textwidth]{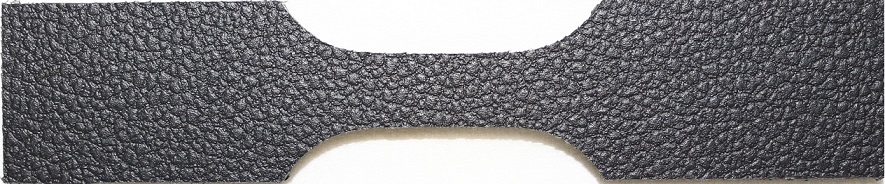}
\caption{Pattern of the UT specimen component cutout}
\label{fig:HSK_specimen_pattern}
\end{figure}

\noindent The elastic modulus, the yield stress as well as the hardening behavior from the component UT cutouts are almost identical compared to the calibration tests with the specimens from plates. However, the fracture strains of the component UT specimens are significantly lower, even though the same specimen geometry, boundary and loading conditions were applied as for the UT specimens from plates. Possible reasons for the deviation in fracture strain are:

\begin{itemize}
\item The component specimens have a pattern on one side, which could contribute to an earlier crack initiation due to initial indentations, Figure~\ref{fig:HSK_specimen_pattern}. The specimens from plates were smooth on both sides.
\item The UT specimens from the component have a thickness of $\SI{2.9}{mm}$ compared to the $\SI{2.3}{mm}$ thick specimens from the plates. Therefore, the aforementioned skin-core effect could play a role, explaining the slightly lower strength in the component specimens.
\item The smaller sprue gate during the manufacturing of series components could lead potentially more damage for the polymer chains due to higher shear stresses in the manufacturing process.
\end{itemize}

\subsection{Glove box flap component test}

\begin{figure}[H]
\centering
\graphicspath{{images/PT/}}
\centering
\def\svgwidth{0.6\textwidth}
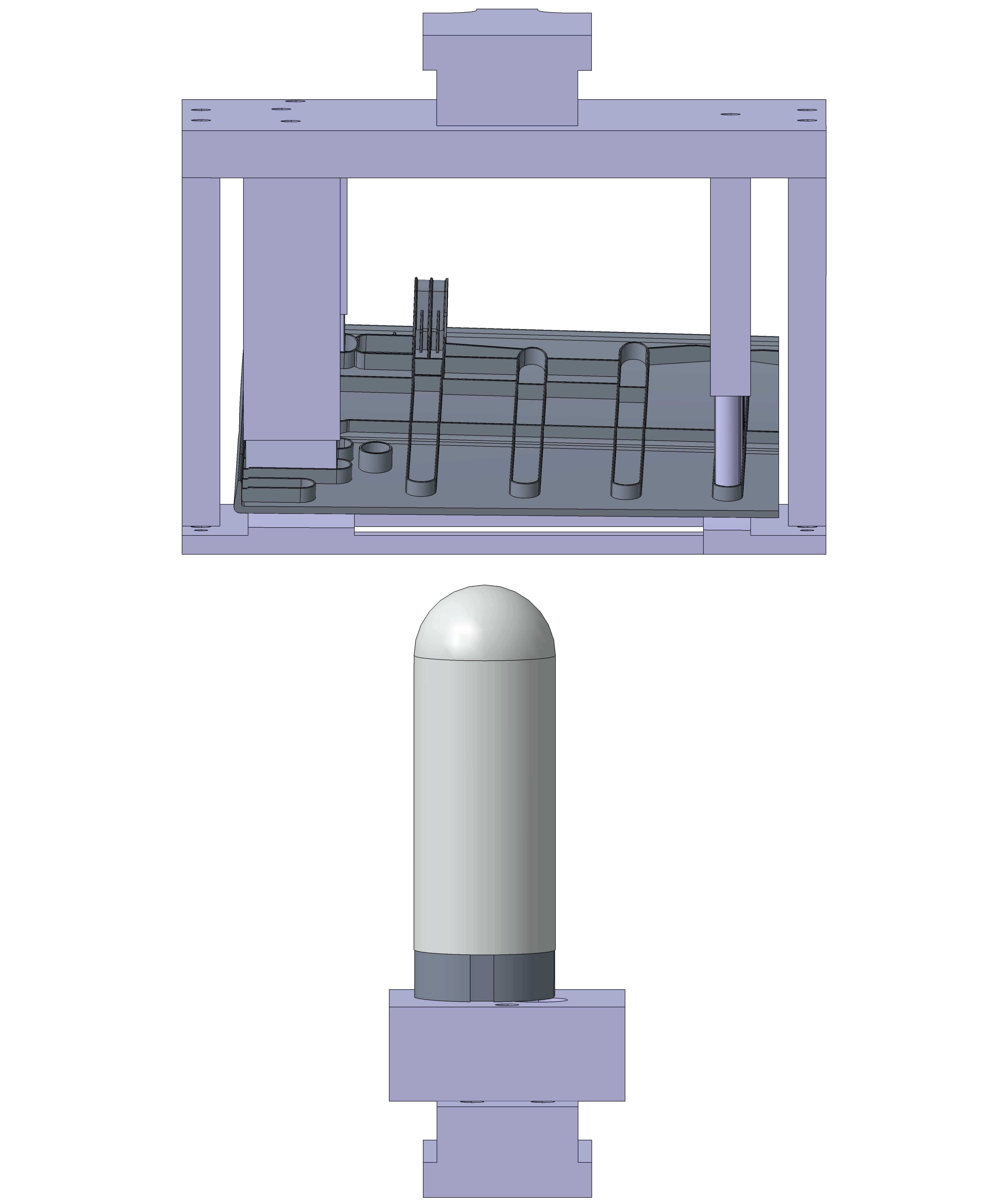
\caption{Experimental set-up of the validation tests}
\label{fig:validation_test_setup}
\end{figure}

\noindent In Figure~\ref{fig:validation_test_setup}, the experimental set-up of the component test is shown. For clamping, a custom holder was designed and built. It allows for the punch to penetrate the component in a defined position while the fracture initiation and propagation is captured with an optical measurement system. The punch has a diameter of $\SI{60}{mm}$ and is made from polytetrafluoroethylene (PTFE) to achieve low friction ($<$0.04 \cite{Walker2016}) between the punch and the component. The force signal is recorded using a $\SI{250}{kN}$ load cell and the displacements are recorded by a Zwick/Roell testing machine. In the experiment, the crosshead displacement was compared to optical measurements of the punch displacement with digital image correlation. Both measurement techniques gave the same displacement values. To guarantee precise measurements even when testing with a temperature chamber, the crosshead displacements are used for all experiments. For the four bolts between the upper and lower clamp plates, a tightening torque of $\SI{15}{Nm}$ each is applied.  \\

%In contrast, Clausen et. al~\cite{Clausen2011} validated their material model with a bending test using a punch made from steel (coefficient of friction between the plate and steel parts was set to 0.1). \\

\noindent The tests were performed at either room temperature (RT), or for testing at elevated and low temperatures, inside a temperature chamber. RT is defined as $\SI{23}{^\circ \text{C}}$. For the tests other than RT, the ambient temperature inside the chamber, the temperature of the clamps and the specimen temperature were controlled and recorded. Instead of measuring the actual specimen that was being tested, a dummy specimen that was close to the tested specimen and that had been placed inside the chamber for the same duration was measured. The ambient temperature, the temperature of the tools and the (dummy) specimen temperature were within the tolerance of $\pm \SI{2}{^\circ \text{C}}$ from the start until the end of the test. After heating/cooling the temperature chamber to its target temperature, each individual specimen was preheated/cooled inside the temperature chamber for a time period of 20 minutes prior to testing. \\

\begin{figure}[H]
\centering
\graphicspath{{images/}}
\def\svgwidth{1.0\textwidth}
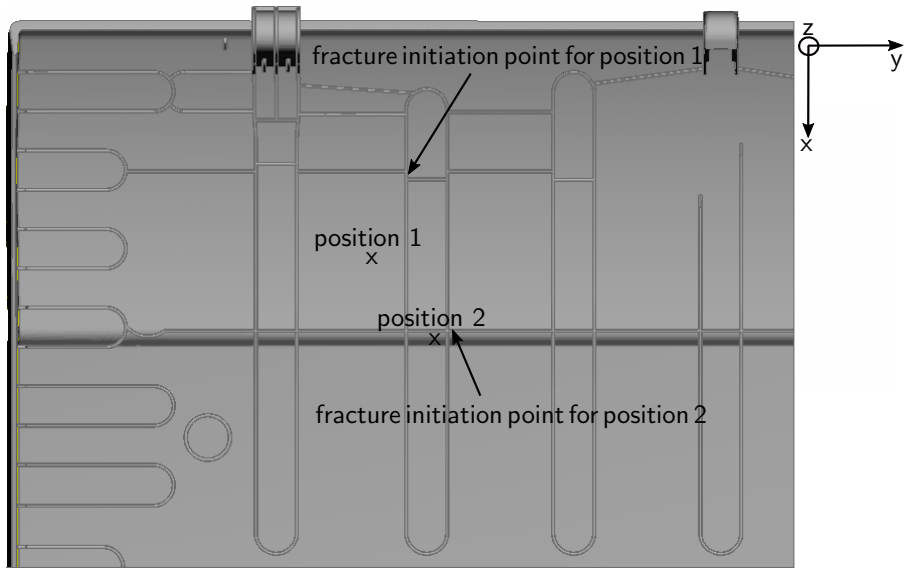
\caption{Punch positions on the glove box flap component}
\label{fig:punch_positions_component}
\end{figure}

\noindent Two punch positions were investigated, as shown in Figure~\ref{fig:punch_positions_component}. It should be noted that the punch is moving in positive z-direction, i.e. coming from the side that is visible to the passenger. For position 2, the fracture initiation point is close to the punch position. In comparison, for position 1, the fracture occurs further away from the first contact point. \\

\begin{figure}[H]
\centering
\graphicspath{{images/boundary/}}
\def\svgwidth{1.0\textwidth}
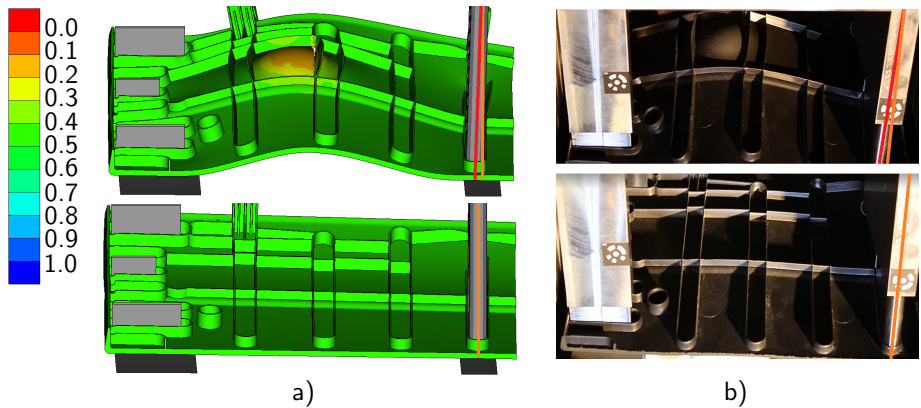
\caption{Slipping/rotation of clamping system}
\label{fig:boundary_slipping}
\end{figure}

\noindent During the experiments, a tilt of the right clamp was observed, see Figure~\ref{fig:boundary_slipping}b. The FE model is shown in Figure~\ref{fig:boundary_slipping}a, colors indicate the accumulated damage of each element. The movement of the right clamp allowed the component to undergo slightly greater deformation. In order to capture this phenomenon, the right clamp in the FE model was given a rotational degree of freedom at the upper right corner, where it is bolted to the upper plate of the holder. This allows the lower right clamp to slip against the lower clamp and account for the slippage in the experiments. The orange line marks the orientation of the right clamp in the undeformed state. The slight offset due to the slippage is visualized with the red line, marking the orientation of the right clamp in the deformed state at the onset of fracture.  \\

\begin{figure}[H]
\centering
\includegraphics[width=\textwidth]{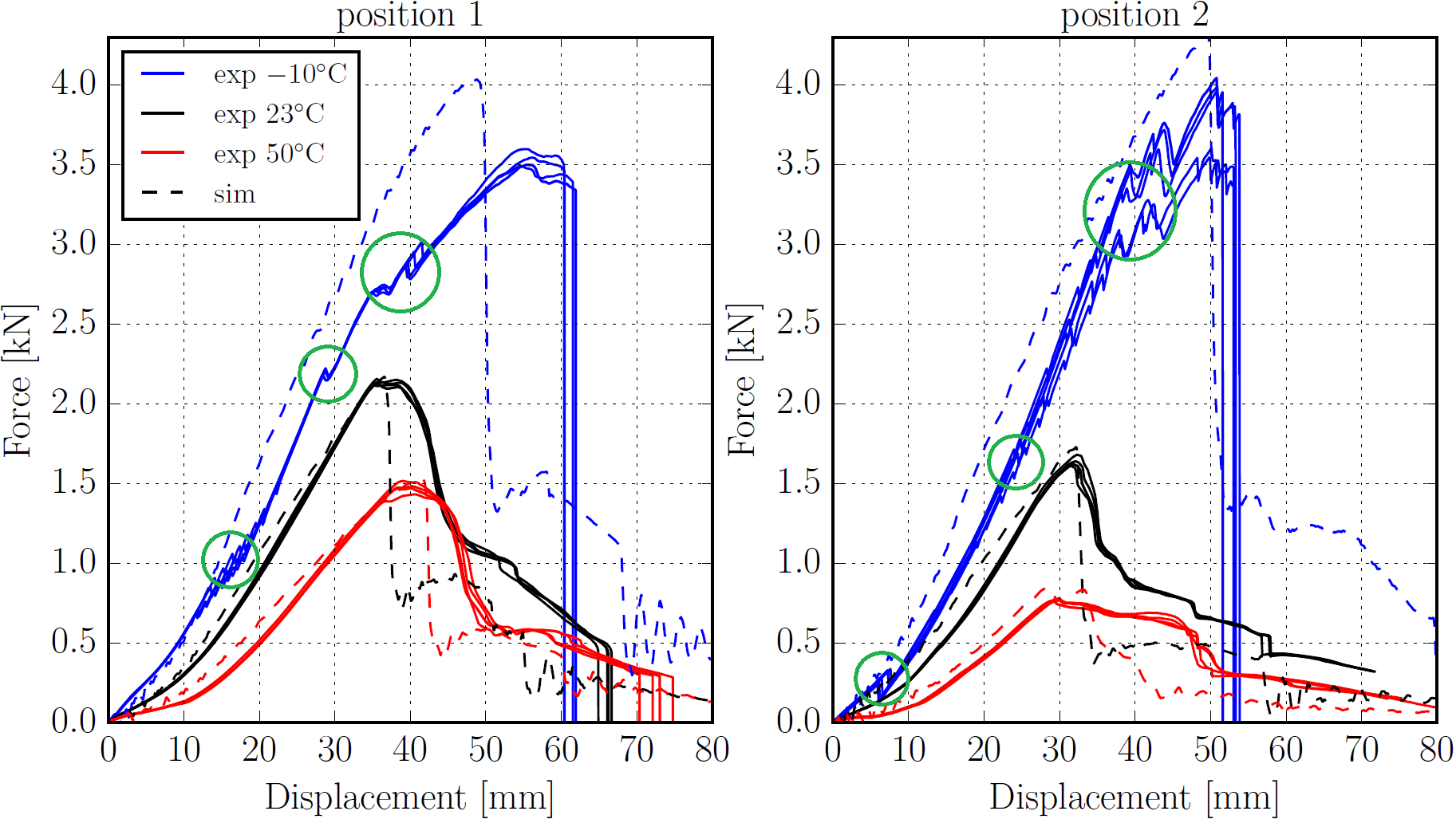}
\caption{Force-displacement curves of the component tests and corresponding simulations for both punch positions}
\label{fig:PTForceDispl_exp_sim_Pos1and2}
\end{figure}

\begin{figure}[H]
\centering
\graphicspath{{images/boundary/}}
\def\svgwidth{1.0\textwidth}
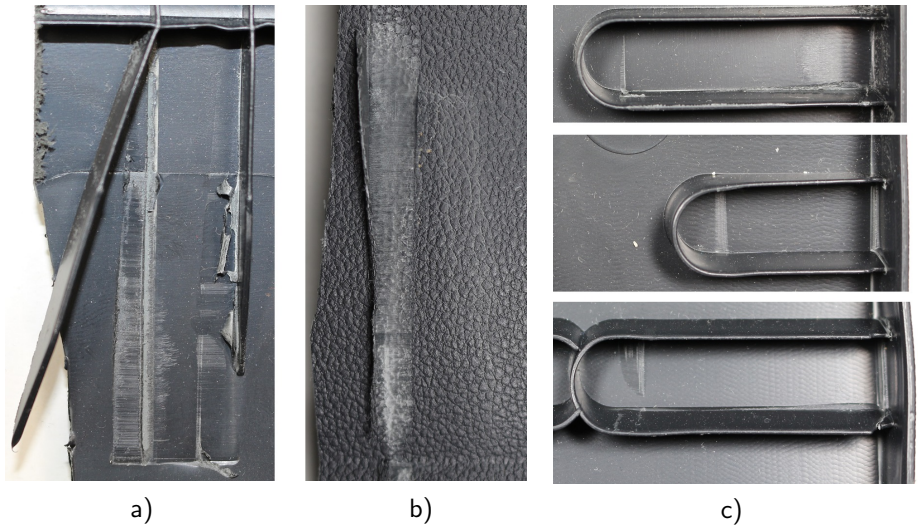
\caption{Fracture at the ribs next to the clamps at $\SI{-10}{^\circ C}$ for punch position 2 \newline \qquad a) Broken rib on the right clamp \newline \qquad b) Damaged surface on the other side of a) \newline \qquad c) Material damage on the left clamp}
\label{fig:damaged_component}
\end{figure}

\noindent In Figure~\ref{fig:PTForceDispl_exp_sim_Pos1and2}, the experimental results of the validation tests at $\SI{-10}{^\circ C}$, $\SI{23}{^\circ C}$ and $\SI{50}{^\circ C}$ are presented for punch positions 1 and 2. Generally, the force slope and the maximum force increase at decreasing temperature while the fracture displacement decreases. At $\SI{-10}{^\circ C}$, fracture occurs fairly early during the experiment at the clamps, seen by the small drops in the force-displacement curve far before reaching the force maximum. The intermediate drops in the blue force-displacement curves ($\SI{-10}{^\circ C}$) marked with circles outline crack initiations in the experiments. As seen from the damaged part from punch position 2, Figure~\ref{fig:damaged_component}, the early crack initiations must have occured at the ribs next to the clamps, which broke off during the experiment. Analogously, for punch position 1 at $\SI{-10}{^\circ C}$, the same rib broke, causing the drops in the force-displacement curves. During the experiments at $\SI{23}{^\circ C}$ and $\SI{50}{^\circ C}$, the first fracture occurs at the central ribs as indicated in Figure~\ref{fig:punch_positions_component}, at the force maximum. Overall, in both positions and at all three ambient temperatures, a good test repeatability was achieved.

\begin{figure}[H]
\centering
\includegraphics[width=\textwidth]{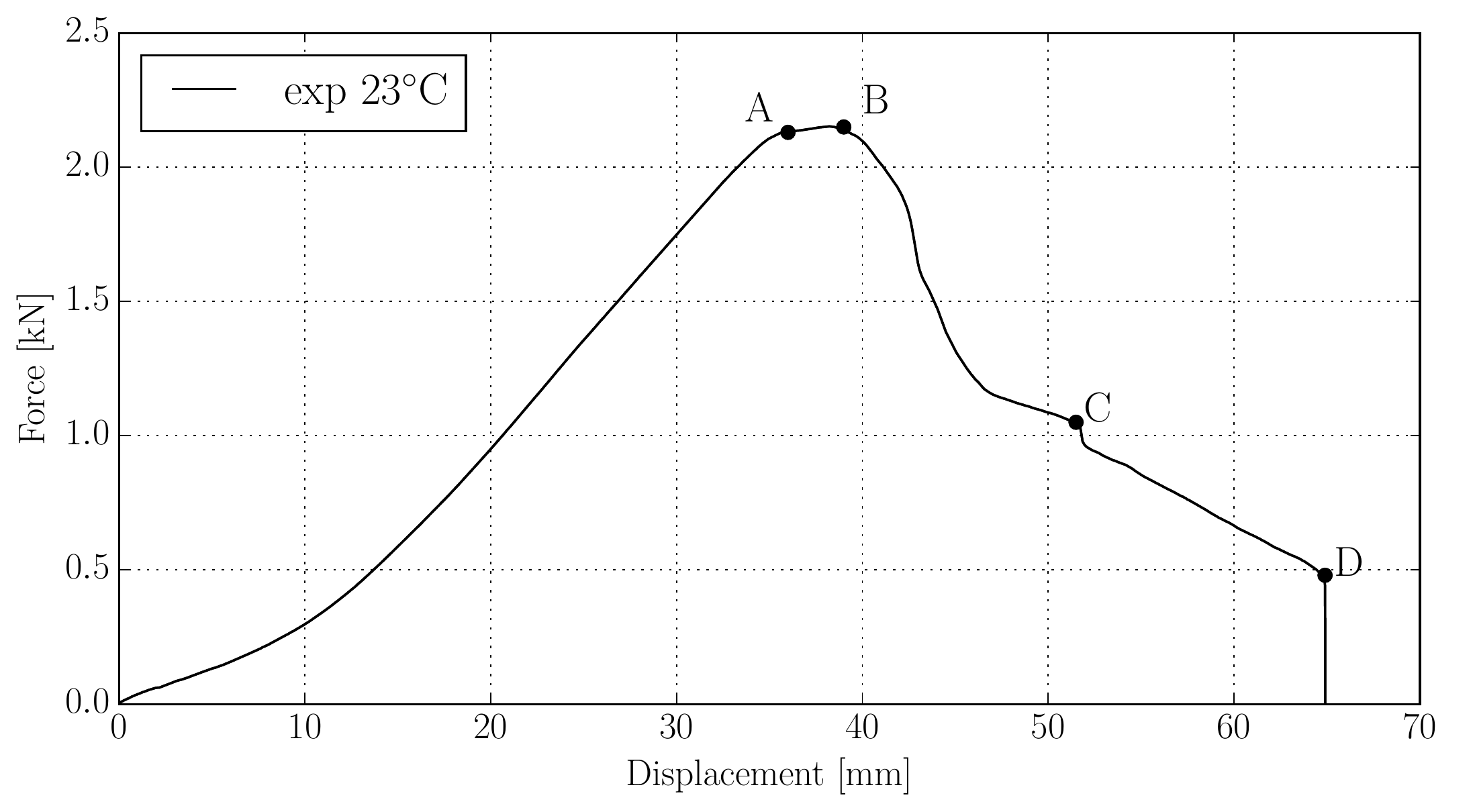}
\caption{Force-displacement curve of the component test with punch position 1 at $\SI{23}{^\circ C}$}
\label{fig:PTForceDisplPos1RTCrackGrowth}
\end{figure}

\begin{figure}[H]
\centering
\graphicspath{{images/PT/Cracks/}}
\def\svgwidth{1.0\textwidth}
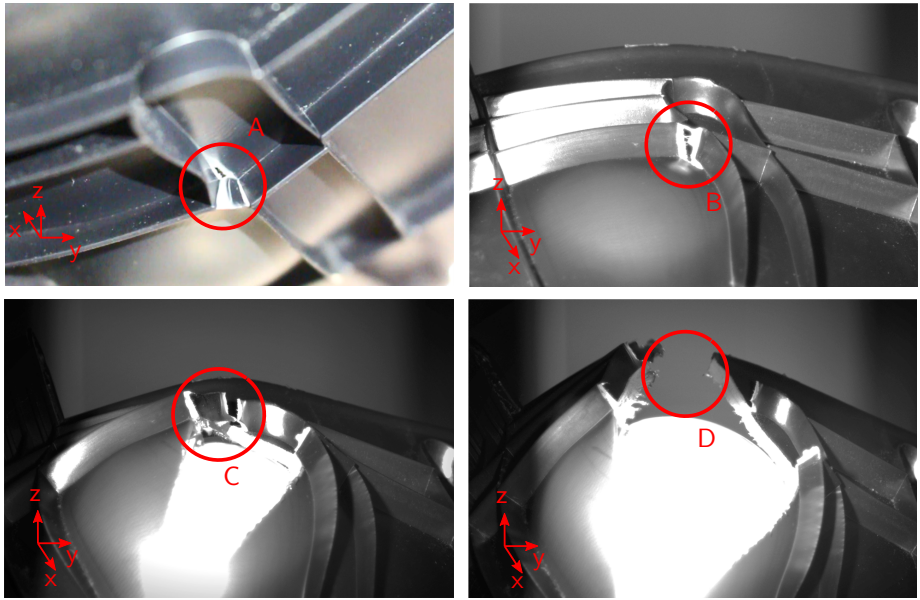
\caption{Fracture propagation of the component test with punch position 1 at $\SI{23}{^\circ C}$}
\label{fig:PTPos1RTCrackGrowth}
\end{figure}

\begin{figure}[H]
\centering
\graphicspath{{images/boundary/}}
\def\svgwidth{1.0\textwidth}
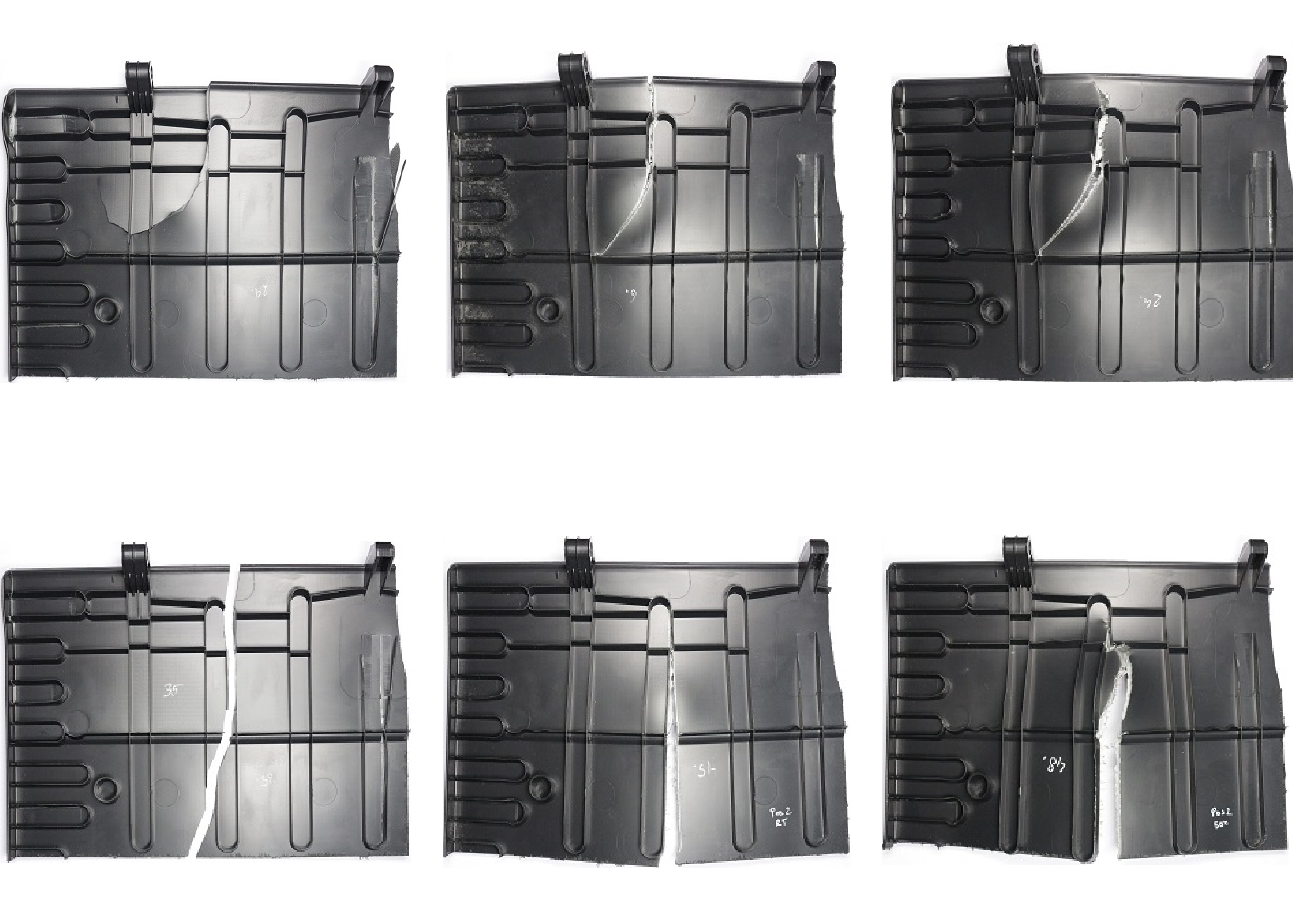
\caption{Crack patterns in the test series of the component for both punch positions at the ambient temperatures $\SI{-10}{^\circ C}$, $\SI{23}{^\circ C}$ and $\SI{50}{^\circ C}$}
\label{fig:crack_pattern}
\end{figure}

\noindent In order to outline the crack growth during the RT tests with punch position 1, the force-displacement curve is shown in Figure~\ref{fig:PTForceDisplPos1RTCrackGrowth}, indicating points where distinctive fracture changes occur in the glove box flap during the test. The corresponding fracture frames are shown in Figure~\ref{fig:PTPos1RTCrackGrowth}. First, fracture occurs in point A, where the inner edge of the rib starts to open. The crack propagates along the edge of the diagonal rib, see point B. After the rib breaks completely, the crack quickly grows along both directions perpendicular to the ribs. A force plateau is reached after the crack is stopped at an additional rib, point C, before the second rib breaks resulting in complete fracture, point D. For the other ambient temperature tests with punch position 1 the first fracture occurs at the same location. In position 2, initial fracture starts at the rib close to the punch position, Figure~\ref{fig:punch_positions_component}, for all three ambient temperatures, growing further along the longitudinal rib. An overview on the different fracture progressions at the two impact positions at the ambient temperatures $\SI{-10}{^\circ C}$, $\SI{23}{^\circ C}$ and $\SI{50}{^\circ C}$ is given in Figure~\ref{fig:crack_pattern}. 

\section{Model validation} \label{Ch:Validation}

\begin{figure}[H]
\centering
\graphicspath{{images/}}
\def\svgwidth{1.0\textwidth}
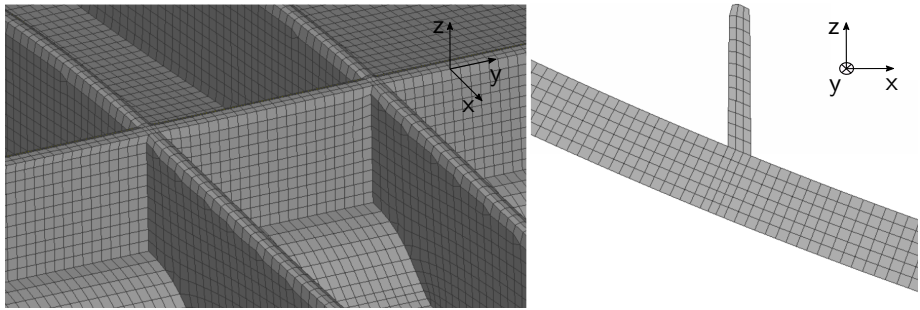
\caption{FE volume element simulation model}
\label{fig:sim_model_hexas}
\end{figure}

\noindent In the simulation model, a fine volume discretization of $\SI{0.5}{mm}$ was used, leading to 1.2 million volume elements for the glove box flap component. The FE model is composed of (mostly) hexahedral elements, where at least 3 elements over the thickness have been used, see Figure~\ref{fig:sim_model_hexas}. The force-displacement curves from the experiments and simulations are compared as well as the crack paths. Therefore, the contact force at the punch and the punch displacement are measured in the simulation model.

\subsection{UT test} \label{Ch:Validation_UT}

\noindent As mentioned in Section~\ref{Sec:experimental_work}, the fracture properties (i.e.~the fracture strains) of the component differ from what was found in the plates used for material model calibration. To account for the smaller fracture strains in the components, a scaling factor $f_\text{f}$ is introduced in the numerical simulations. The purpose of this parameter is to scale down the achievable fracture strain $\bar{\varepsilon}_\text{f,scaled}$ of the elements in order to reach a FRAC-value of 1.0 at the point where the UT tests of the specimens extracted from components fracture, see Figure~\ref{fig:UTStressStrainDINIT}:

\begin{equation} \label{eq:scaled_fracture_value}
\bar{\varepsilon}_\text{f,scaled} = f_\text{f} \, \bar{\varepsilon}_\text{f}
\end{equation}

\noindent All other parameters of the material model (elasticity, plasticity, hardening, softening, etc.) are not influenced by this measure. This way, the simulation model curves originally representing the fracture behavior from plates can be adjusted to now represent the component behavior. The UT specimens extracted from the component were tested at the same main supporting temperatures as used for model calibration. Therefore, the scaling factor $f_\text{f}$ can be easily determined for these temperatures, see Table~\ref{tab:scaling_fracture_values}. For $\SI{20}{^\circ \text{C}}$, the difference in fracture strain between the component and the plates is over 50\%. As outlined by the thickness sensitivity study in Section~\ref{Ch:material_description}, a majority of this loss happens due to the thickness effect. The remaining difference could be explained by the pattern of the component specimens and the difference in sprue gate areas. For the other ambient temperatures, $\SI{-35}{^\circ \text{C}}$ and $\SI{90}{^\circ \text{C}}$, the thickness influence is significantly smaller, indicating that the properties of the skin and the core have more closely converged.

\begin{table}[H]
\centering
\caption{Scaling factors for main supporting ambient temperatures}
\centering
\begin{tabular}{ccccc} \toprule
Temperature & $T$ & $\SI{-35}{^\circ \text{C}}$ & $\SI{20}{^\circ \text{C}}$ & $\SI{90}{^\circ \text{C}}$  \\ \midrule
Scaling factor & $f_\text{f}$ & 0.80 & 0.45 & 0.80 \\ \bottomrule
\end{tabular}
\label{tab:scaling_fracture_values}
\end{table}

\subsection{Glove box flap component test} \label{Ch:Validation_component}

\begin{figure}[H]
\centering
\graphicspath{{images/solidshell/}}
\def\svgwidth{1.0\textwidth}
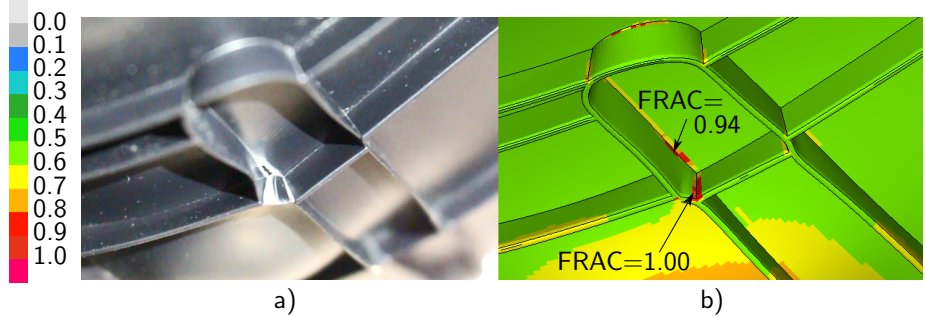
\caption{Comparison of initial fracture in the component experiment (a) and the simulation (b) at position 1 and $\SI{23}{^\circ C}$}
\label{fig:validation_component_exp_sim_pos1}
\end{figure}

\noindent In Figure~\ref{fig:validation_component_exp_sim_pos1}, the initial fracture in the component experiment at $\SI{23}{^\circ \text{C}}$ is shown for punch position 1, compared to the simulation model after $\SI{15}{mm}$ punch displacement. At this point, first fracture occurs in the experiments. At the same time, the first elements in the simulation have reached a FRAC-value of 1.0 in the rib and were eliminated from the simulation. Even though the punch penetrates the component at a position away from the rib, see Figure~\ref{fig:punch_positions_component}, first fracture occurs in the rib. This is well captured by the material model when using a fine volume discretization. At the other ambient temperatures $\SI{-10}{^\circ \text{C}}$ and $\SI{50}{^\circ \text{C}}$ initial fracture occurs at the same rib.

\begin{figure}[H]
\centering
\graphicspath{{images/solidshell/}}
\def\svgwidth{1.0\textwidth}
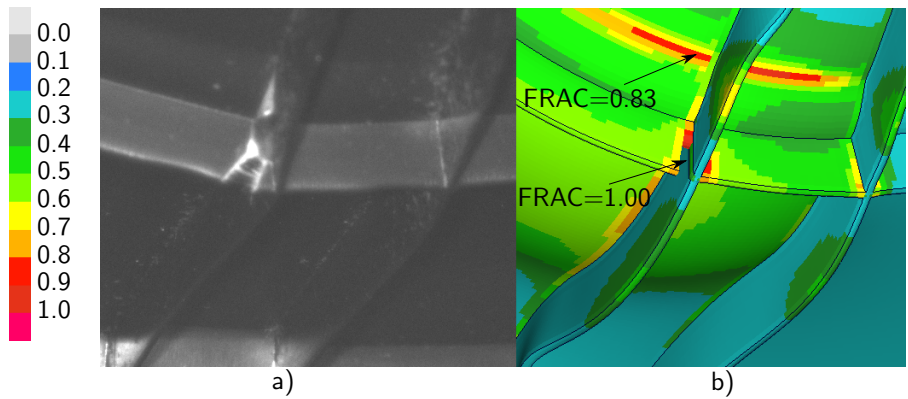
\caption{Comparison of initial fracture in the component experiment (a) and the simulation (b) at punch position 2 and $\SI{-10}{^\circ C}$}
\label{fig:validation_component_exp_sim_pos2}
\end{figure}

\noindent For punch position 2, the onset of fracture is shown in Figure~\ref{fig:validation_component_exp_sim_pos2} for the experiments and simulations at $\SI{-10}{^\circ \text{C}}$. The initial fracture location is well predicted by the simulation. Again, fracture is initiated in a rib. From there, the initial crack starts growing parallel to the rib until global fracture for all three ambient temperatures $\SI{-10}{^\circ \text{C}}$, $\SI{23}{^\circ \text{C}}$ and $\SI{50}{^\circ \text{C}}$. Simulation and experiment correlate well with regard to initial fracture, but also fracture propagation. However, in order to capture the post-fracture behavior precisely, an additional model is required. This can be seen in the deviation of experiment and simulation in Figure~\ref{fig:PTForceDispl_exp_sim_Pos1and2} after the force maximum has been reached. \\

\subsection{Shell vs. volume discretization} \label{Ch:Validation_discretization}

\begin{figure}[H]
\centering
\graphicspath{{images/solidshell/}}
\def\svgwidth{1.0\textwidth}
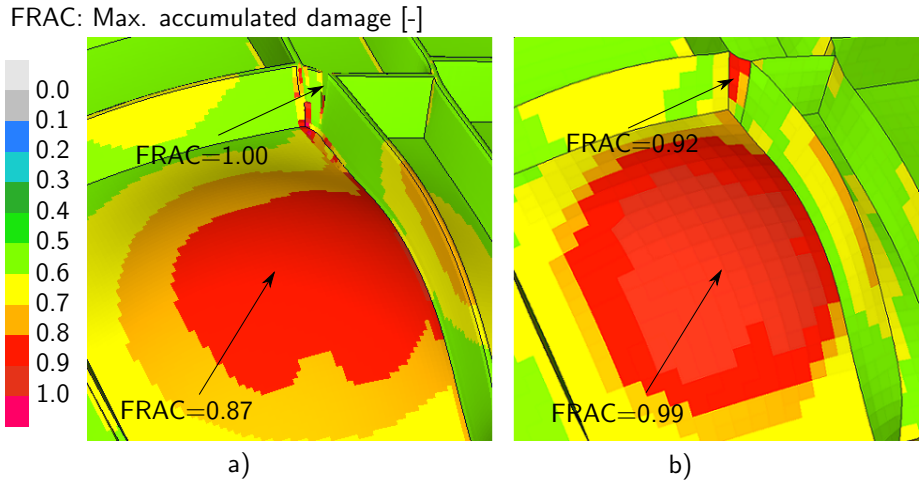
\caption{Comparison of $\SI{0.5}{mm}$ volume (a) and $\SI{2.0}{mm}$ shell (b) discretization, punch position 1 at $\SI{23}{^\circ \text{C}}$}
\label{fig:validation_component_exp_sim}
\end{figure}

\noindent In Figure~\ref{fig:validation_component_exp_sim}, the $\SI{0.5}{mm}$ volume element discretization is compared to the elementation using $\SI{2}{mm}$ shell elements. Even with the coarse shell elementation, the fracture location in the rib is detected as critical, however, because the strains are not localizing enough in the sharp edge, the first cracks occur at the position close to the punch, rather than in the edge as in the experiment. A finer shell element size may help to identify the rib as position of initial fracture. \\

\section{Conclusions and recommendations}

\noindent In this work, the temperature-dependent material model proposed in \cite{Degenhardt2019} is validated using glove box flap component punch tests. Therefore, a custom holder was designed and built. In the experiments, two punch positions at three different ambient temperatures were investigated. \\

\noindent In order to verify the material properties of the component, UT specimens were extracted and compared to specimens milled from plates, which were used for the material model calibration. The preliminary UT tests outlined that the component UT specimens show a smaller fracture displacement than the UT specimens from plates, varying with ambient temperature. The smaller fracture displacement is accounted for in the simulation model by introducing a scaling factor for the fracture strain applied for all elements in the initial state. This way, the simulation model curves originally representing the material behavior from plates can be adjusted to represent the component behavior. With this consideration, the material model captures the stiffness, the hardening, the force-maximum and the fracture strain for all ambient temperature component experiments with very good accuracy. For $\SI{-10}{^\circ \text{C}}$, cracks occur at the clamps shortly after loading, however, they do not change the global fracture behavior. Generally, the temperature-dependent material model proves to be a good model to represent the material behavior of series production components. \\

\noindent With shell elements of $\SI{2}{mm}$ element length or greater, the stresses in sharp edges may be underestimated. When using fine volume discretization, the material model captures the material behavior even at sharp edges with high notch stresses. It is then possible to predict the behavior up to the point of initial fracture. \\

\section*{References}

%%%\bibliography{lit}

\end{document}